\documentstyle[11pt,paspconf,epsf]{article}

\begin{document}

\title{Oceanography of Accreting Neutron Stars: Non-Radial
Oscillations and Periodic X-Ray Variability}

\author{Lars Bildsten, Andrew Cumming, \& Greg Ushomirsky}
\affil{Department of Physics, Department of Astronomy, 
  University of California, Berkeley, CA 94720}

\author{Curt Cutler} 
\affil{Max-Planck-Institut fuer Gravitationsphysik,
Albert-Einstein-Institut, Sclaatzweg 1, 14473 Potsdam, Germany}

\begin{abstract}

Observations of quasi-periodic oscillations (QPOs) in the luminosity
from many accreting neutron stars (NS) have led us to investigate a
source of periodicity prevalent in other stars: non-radial
oscillations.  After summarizing the structure of the atmosphere and
ocean of an accreting NS, we discuss the various low $l$ g-modes with
frequencies in the 1-100 Hz range. Successful identification of a
non-radial mode with an observed frequency would yield new
information about the thermal and compositional makeup of the NS, as
well as its radius. We close by discussing how rapid rotation changes
the g-mode  frequencies.

\end{abstract}


\section{Introduction and Summary} 

The liquid core, solid crust and overlying ocean give a NS a very rich
non-radial oscillation spectrum, with p-modes in the 10kHz range and a
variety of lower frequency modes from the toroidal and spheroidal
displacements of the crust and surface layers (see McDermott et
al. 1988 for an overview). Non-radial g-mode oscillations have been
studied extensively for isolated radio pulsars, where the atmosphere
and ocean are quiescent. These studies can be separated by the sources
of buoyancy: (1) entropy gradients (McDermott, Van Horn \& Scholl
1983, and McDermott et al. 1988), (2) density discontinuities (Finn
1987, McDermott 1990, Strohmayer 1993) and, (3) mean molecular weight
gradients due to $\beta$ equilibrium in the core (Reisenegger \&
Goldreich 1992). To date, none of these modes has been securely
identified with any particular  radio pulsar phenomenon.

There are also many NSs accreting in binary systems, with internal
structures quite different from that of a radio pulsar. There is a
rich spectrum of QPO's in the luminosity from these accreting NSs
which has motivated much of our work. Utilizing EXOSAT data on the
highest accretion rate objects ($\dot M> 10^{-10} M_\odot \ {\rm
yr^{-1}}$), Hasinger \& van der Klis (1989) found that the objects
split into two separate classes. Six objects trace out all or part of
a ``Z'' in an X-ray color-color diagram and exhibit time dependent
behavior that correlates with the position along the Z.  These ``Z''
sources have QPO's in the 15-50 Hz and 5-7 Hz range with up to 10\%
modulation. The other objects fall into separated regions of the
color-color diagram and do not show similar QPO phenomenology at less
than 100 Hz. These ``Atoll'' sources accrete at lower rates than the Z
sources and exhibit Type I X-ray bursts resulting from the unstable
ignition of the accumulated hydrogen and helium (see Bildsten 1998 for
a recent review). The Rossi X-Ray Timing Explorer (RXTE) has detected
coherent oscillations during these Type I bursts (see Strohmayer et
al. 1996 for an example). RXTE also detected two drifting kHz QPOs in
the persistent emission which are separated by a constant frequency
that is identical to that seen in the burst. This naturally leads to a
beat frequency model, where the difference frequency is presumed to be
the NS spin. The origin of the upper frequency differs in
the models (see van der Klis 1997 for a summary), and usually involves
the Kepler frequency at some point in the accretion flow (Miller et
al. 1996, Kaaret et al. 1997, Zhang et al. 1997).

It might be that all of the $<100$ Hz and kHz QPOs can be accomodated
within the existing models (see van der Klis 1995, 1997) that use the
accretion disk and/or spherical flow to generate the periodic
phenomena. However, it is also important to pursue non-radial
oscillations as a possible source of some of these periodicities
(McDermott \& Taam 1987, Bildsten \& Cutler 1995 (BC95), Bildsten,
Ushomirsky \& Cutler 1996 (BUC), Strohmayer \& Lee 1996 (SL96),
Bildsten \& Cumming 1997 (BC97)). These offer the distinct advantage
of having well understood dispersion relations (even when rapidly
rotating) and frequencies (and frequency derivatives) that depend on
the underlying NS structure.  This makes it a challenge to
successfully identify an observed QPO with a non-radial
oscillation. However, for the same reason, this hypothesis offers much
potential, as the successful identification of a non-radial pulsation
with an observed frequency would tell us about the NS radius and
internal structure.

\section{The Outer Layers of an Accreting Neutron Star}

 In most cases where stellar pulsations are studied, the underlying
stellar model is reasonably well understood and constrained via other
observations. The situation is quite different in an accreting NS,
where the conditions in the outer layers can change on timescales of
hours and the deep internal structure depends on the accretion
history. We just roughly sketch the situation here, more details can
be found in BC95, SL96, and BC97.

The NSs in mass-transferring binaries accrete the hydrogen (H) and
helium (He) rich matter from the surfaces of their companions and
undergo unstable H/He burning when $\dot M < 10^{-8} M_\odot \ {\rm
yr^{-1}}$. This typically occurs at $\rho < 10^6 \ {\rm g \ cm^{-3}}$,
and within a few hours to days upon arrival on the star.  The very
high temperatures reached ($T> 10^9 {\rm \ K}$) during the thermal
instability produce elements at and beyond the iron group.  The
isotopic mixture from this burning is still not well known, though
everyone agrees that a substantial amount of H (the residual mass
fraction is $X_r\sim 0.1-0.5$) remains unburned (see Bildsten 1998 for
a review). This matter accumulates on the NS and forms a
relativistically degenerate ocean. The hydrogen is eventually depleted
due to electron captures at $\rho \approx 10^7 \ {\rm g \ cm^{-3}}$,
leading to a density discontinuity. The material crystallizes and
forms the NS crust at $\rho \sim 10^8-10^9 \ {\rm g \ cm^{-3}}$. There
is no evidence for magnetic fields on these NSs, and most arguments
about the nature of Type I X-ray bursts limit $B < 10^9 \ {\rm G}$,
weak enough to not affect the ocean g-modes (BC95).

\section {The Adiabatic Non-Radial Oscillations in the Deep Ocean} 

The g-modes cannot penetrate into the crust, as the restoring force
from the finite shear modulus effectively excludes them (McDermott et
al. 1988, BC95), and so they reside in the relatively thin ocean
(thickness is $\approx 10^4 \ {\rm cm} \ll R$). The low $l$ modes are
in the shallow water wave limit and so $\omega^2\propto k^2$, where
$k=(l(l+1))^{1/2}/R$ is the transverse wavenumber for a slowly
rotating star. Prior work has been done on g-modes in the upper
atmosphere. McDermott \& Taam (1987) calculated g-modes of a bursting
atmosphere. SL96 calculated the non-adiabatic mode structure for
atmospheres accreting and burning in steady-state and found that
g-modes may be excited by the $\epsilon$ mechanism when
$\dot{M}<10^{-10}M_\odot \ {\rm yr^{-1}}$. However, steady-state
burning does not occur at these low $\dot M$'s, and there have yet to
be realistic calculations for the time dependent NS atmosphere.

We have focused on the adiabatic mode structure in the deep ocean
underneath the H/He burning where the thermal time (hours to
days) is much longer than the mode period. The different sources of
buoyancy yield a rich spectrum of g-modes. The abrupt rise in density
associated with the hydrogen electron capture boundary layer supports
a density discontinuity mode of frequency (BC97)
\begin{equation}\label{eq:fd}
f_d\approx 35\ {\rm Hz}
\left({X_r}\over {0.1}\right)^{1/2}
\left({10 \ {\rm km}}\over R\right)
\left( {l(l+1)\over 2} \right)^{1/2}, 
\end{equation}
and the internal buoyancy due to the composition gradient within the
electron capture boundary layer creates a new spectrum of modes which
are ``trapped'' (BC97). Almost all of the mode energy is confined to
the boundary layer and a WKB estimate of the mode frequency gives
(BC97)
\begin{equation}\label{eq:fn}
f_{tr}\approx \frac{8.5 \ {\rm Hz}}{n}
\left({X_r}\over {0.1}\right)^{1/2}
\left({10 \ {\rm km}}\over R\right)
\left( {l(l+1)\over 2}\right)^{1/2},
\end{equation}
where $n$ is the number of nodes in the boundary layer and we have
omitted the weak dependence on the accretion rate.  There is also a
set of thermal g-modes (BC95), which we have recently found (BC97) are
separated by the density discontinuity, and confined to either the
upper or lower parts of the ocean.  The isothermal ocean is
relativistically degenerate, so that the dimensionless density
contrast due to the entropy gradient is small and $\propto
k_BT/E_F$, allowing BC95 to obtain the exact analytic formula
\begin{equation}
f_{th}=6.3 \ {\rm Hz} \left[{l(l+1)\over 2}{T_8\over\Theta_n}{16\over\mu_i}
\right]^{1/2}\left({10 \ {\rm km}\over R}\right),
\end{equation}
where $\Theta_n=1+(3n\pi /2\ln(\rho_b/\rho_t))^2$ corrects for $n$ and
the density contrast between the top and bottom of the ocean, 
$T_8=T/10^8 \ {\rm K}$  and $\mu_i$ is the ion mean molecular weight. 

\section{The Role of Rapid Rotation and the Future} 

\begin{figure} 
\begin{center}
%\plotfiddle{rotate.ps}{2.8in}{0}{37}{37}{0}{0}
\plottwo{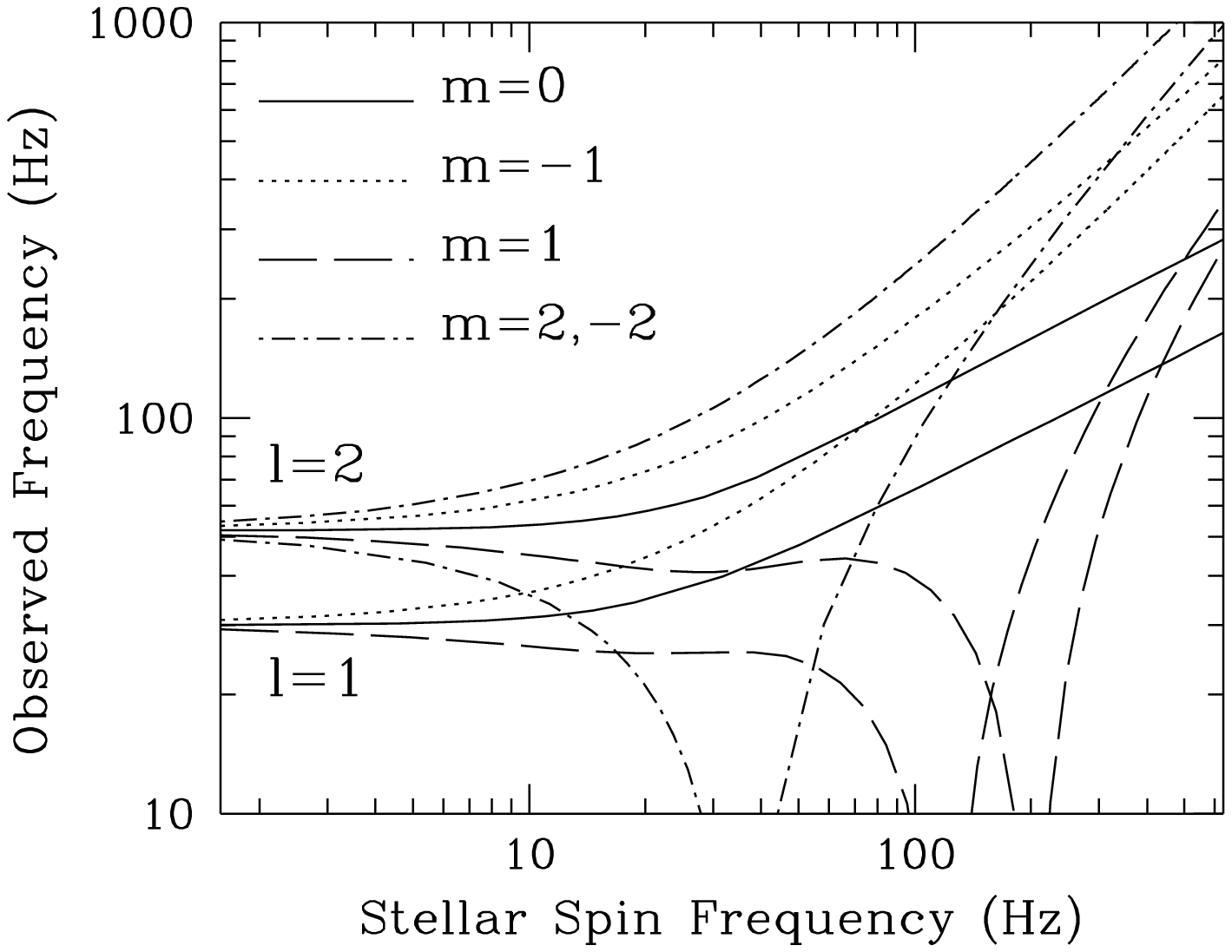}{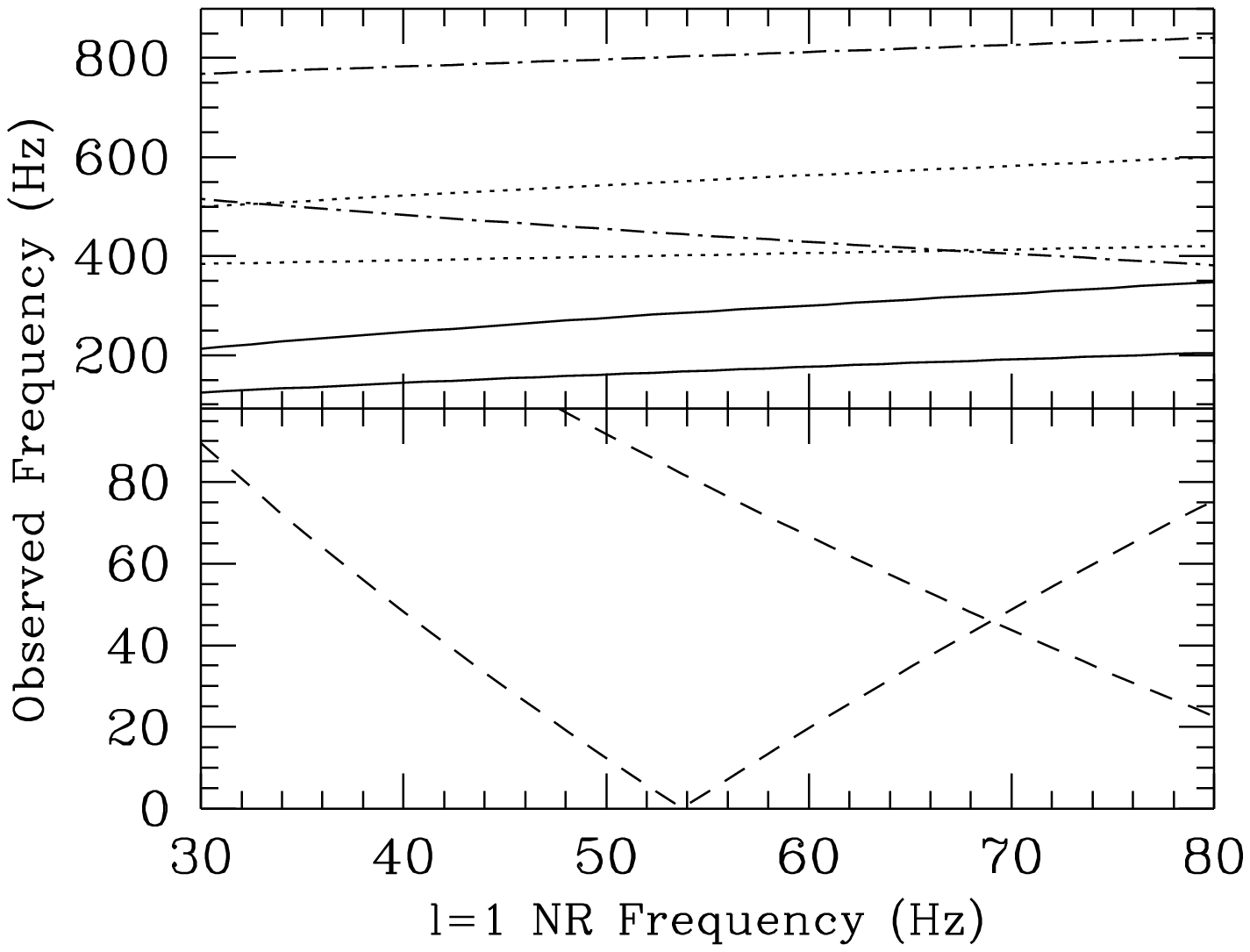}
\caption{{\bf Left Panel: } The frequencies of a mode which is at 30
Hz ($l=1$) for a non-rotating star as a function of the NS spin
frequency.  This could be a hydrogen electron capture density
discontinuity mode or a thermal mode in the upper burning layers. The
lines are for different values of $m$ and start out, at slow rotation,
being the split modes of a given $l$. {\bf Right Panel: } The
observable frequencies of modes when $f_s=363$ Hz
for different values of $m$ (same as in left panel) as a function of
the changing internal dynamics. The modes in the kHz range show little
dynamic range and are not split by the spin frequency. The
bottom panel shows the retrograde $m=1$ modes that are at low
frequencies and can change rapidly.
\label{fig:rapidrot}}
\end{center}
\end{figure}

We might suspect that these stars are rapidly rotating, as the
prolonged accumulation of material will most likely spin up the star.
The coherent periodicities during Type I X-ray bursts (see \S 1) seem
to indicate 500 Hz spin frequencies, which are much greater than
the g-mode frequencies, but still small compared to the breakup
frequency $\Omega_b\approx (GM/R^3)^{1/2}$ ($\approx 2$ kHz for a $1.4
M_\odot, R=10$ km star). When $\Omega \ll \Omega_b$, the unperturbed
star is spherical and the centrifugal force can be neglected, in which
case the primary difference in the momentum equations is the Coriolis
force. As a result, the g-mode frequencies depart significantly from
the $\omega^2 \propto l(l+1)$ scaling (Papaloizou \& Pringle 1978).

BUC made progress on this problem within what is called the
``traditional approximation'', where the radial and transverse
momentum equations separate and the resulting angular ODE must be
solved to find the angular eigenfunction (no longer just $Y_{lm}$'s)
and transverse eigenvalue $\lambda$ (i.e., we write the transverse
wavenumber as $k^2=\lambda/R^2$) which, for a non-rotating star, is
$l(l+1)$. The radial equations are identical to the non-rotating case,
so that if $\omega_0$ is the eigenfrequency for the $l=1$ mode of a
{\it non-rotating} star, then the oscillation frequency (in the
rotating frame) at arbitrary spin is
$\omega=\omega_0(\lambda/2)^{1/2}$.  These are then transferred into
the observer's inertial frame via $\omega_I=\omega-m\Omega$.  The left
panel of Figure 1 shows the observed frequencies for a mode that is at
30 Hz ($l=1$) in the non-rotating star. It is intriguing to ask how
the observed frequencies change as the internal conditions change (say
the $T$ or $X_r$ in equations 1-3).  The right panel of Figure 1 shows
this for 4U 1728-34 ($f_s=363$ Hz, Strohmayer et al. 1996). There are
frequencies near a kHz, but they do not show the large dynamic range
exhibited by the observed kHz QPOs. The splitting for some of the
prograde modes is nearly constant, but is not equal to the spin
frequency. Also shown (bottom panel on the RHS) are two retrograde
modes that appear at low frequencies, clearly showing a large dynamic
range for only a small change in the non-rotating mode frequency. This
points to the possibility of explaining some of the QPOs seen in the
Atoll and Z sources.

Our work might eventually provide natural explanations for some of the
observed QPO's in accreting NS. However, there is still much to do
theoretically, from understanding how the modes are excited to how
they modulate the luminosity. Observational progress will come by
identifying QPOs with modes of different $l$'s, where the ratio of the
mode frequencies are known.  This is easiest to do if the NSs are
rotating slowly (spin frequency $f_s < 10$ Hz), but can also be
accomodated if the NSs are rapidly rotating, especially for those NS
where the spin frequency was measured during a Type I burst. 

\acknowledgments

This work is supported by NASA grant NAG 5-2819.  G. U. thanks the
Fannie and John Hertz Foundation for fellowship support.


\begin{references}


\noindent
\reference 
Bildsten, L. 1998, to appear in ``The Many Faces of Neutron Stars'',
ed. A. Alpar, L. Buccheri, and J. van Paradijs (Dordrecht: Kluwer), 
astro-ph/9709094

\reference 
Bildsten, L. \& Cumming, A. 1997, to be submitted to Ap. J. (BC97) 

\reference 
Bildsten, L. \& Cutler, C. 1995, \apj, 449, 800 (BC95)

\reference 
Bildsten, L., Ushomirsky, G., \& Cutler, C. 1996, \apj, 460, 827
(BUC)

\reference 
Finn, L. S. 1987, MNRAS, 227, 265

\reference 
Hasinger, G. \& van der Klis, M. 1989, A\&A, 225, 79 

\reference 
Kaaret, P., Ford, E. C. \& Chen, K. 1997, \apj, 480, L27

\reference 
McDermott, P. N. 1990, MNRAS, 245, 508

\reference 
McDermott, P. N. \& Taam, R. E. 1987, \apj, 318, 278

\reference 
McDermott, P. N., Van Horn, H. M., \& Hansen, C.J. 1988, \apj, 325, 725

\reference 
McDermott, P. N., Van Horn, H. M., \& Scholl, J. F. 1983, \apj, 268, 837

\reference 
Miller, M. C., Lamb, F. K., \& Psaltis, D. 1996, submitted to Ap. J. 

\reference 
Papaloizou, J. \& Pringle, J. E. 1978, MNRAS, 182, 423 

\reference
Reisenegger, A. \& Goldreich, P. 1992, \apj, 395, 240 

\reference 
Strohmayer, T. E. 1993, \apj, 417, 273

\reference 
Strohmayer, T. E. \& Lee, U. 1996, \apj, 467, 773 (SL96) 

\reference 
Strohmayer, T. E., et al. 1996, \apj,  469, L9

\reference 
van der Klis, M. 1995, in ``X-Ray
Binaries'', ed. W. H. G. Lewin, J. van Paradijs \& E. P. J. van den
Heuvel (London: Cambridge), p. 252 

\reference 
van der Klis, M. 1997, to appear Proceedings of the Wise Obs. 
25th Anniv. Symposium (astro-ph/9704272)

\reference 
Zhang, W., Strohmayer, T. E., \& Swank, J. H., 1997, \apj, 482, L167 

\end{references}
\end{document}